# Investigating the Role of Socio-organizational Factors in the Information Security Compliance in Organizations

**Ahmed AlKalbani**
School of Business Information Technology and Logistics
RMIT University
Melbourne, Australia
Email: ahmed.al-kalbani@rmit.edu.au

**Hepu Deng**
School of Business Information Technology and Logistics
RMIT University
Melbourne, Australia
Email: hepu.deng@rmit.edu.au

**Booi Kam**
School of Business Information Technology and Logistics
RMIT University
Melbourne, Australia
Email: booi.kam@rmit.edu.au

## Abstract

The increase reliance on information systems has created unprecedented challenges for organizations to protect their critical information from different security threats that have direct consequences on the corporate liability, loss of credibility, and monetary damage. As a result, the security of information has become critical in many organizations. This study investigates the role of socio-organizational factors by drawing the insights from the organizational theory literature in the adoption of information security compliance in organizations. Based on the analysis of the survey data collected from 294 employees, the study indicates management commitment, awareness and training, accountability, technology capability, technology compatibility, processes integration, and audit and monitoring have a significant positive impact on the adoption of information security compliance in organizations. The study contributes to the information security compliance research by exploring the criticality of socio-organizational factors at the organizational level for information security compliance.

## 1 Introduction

Information security compliance refers to the implementation of information security standards and policies for protecting information in organizations (AlKalbani et al. 2014; Von Solms 2005). The adoption of information security compliance ensures that information security mechanisms can work together effectively to protect the critical information in organizations (Appari et al. 2009; Ifinedo 2013). It satisfies the security requirements, thus improving stakeholders' confidence and trust in organizations. As a result, information security compliance is widely considered as an effective approach for ensuring information security in organizations (Herath and Rao 2009).

Several studies have investigated the problem of information security compliance in organizations in recent years. Herath and Rao (2009), for example, investigate the factors related to behaviours, motivations, values and norms that affect employees' intentions to comply with information security compliance in organizations. Siponen et al. (2010) examine the factors related to normative beliefs, threat appraisal, self-efficacy, and visibility that influence employees' intention to comply with information security policies in organizations. Ifinedo (2013) assesses the social influence of changing individual's thoughts, actions, feelings, attitudes, and behaviours on information security compliance in organizations. These studies have focused primarily on understanding employees' attitudes, and behaviour (Herath and Rao 2009) on information security compliance in organizations. There are, however, other socio-organizational aspects that may influence the adoption of information security compliance in organizations. These aspects include information security governance (Smith and Jamieson 2006), legislative requirements (Benabdallah et al. 2002), information security strategies and policies (Smith and Jamieson 2006), and implementation of advanced security technologies (Lambrinoudakis et al. 2003). This shows that there is a need to investigate more social-organizational factors for shaping the adoption of information security compliance in organizations (Bulgurcu et al. 2010; Dhillon and Backhouse 2001).

This study investigates the role of socio-organizational factors by drawing the insights from the organizational theory literature in the adoption of information security compliance in organizations. Based on the analysis of the survey data collected from 294 employees, the study indicates management commitment, awareness and training, accountability, technology capability, technology compatibility, processes integration, and audit and monitoring have a significant positive impact on the adoption of information security compliance for information security in organizations. Theoretically the study contributes to the information security compliance research by exploring the criticality of socio-organizational factors at the organizational level for information security compliance. Practically this study provides organizations with useful guidelines for adopting effective information security compliance for information security.

The rest of this paper is organized as follows. In section 2, the related literature is reviewed, leading to the development of a theoretical foundation for this study in Section 3. A description of the research methodology is followed in Section 4. Subsequently, the findings and their implications of this study are presented in Section 5. Finally in Section 6, the conclusion, the limitations of the study and future research are given.

## 2  Literature Review

Non-compliance to information security standards and policies is one of the main reasons for security breaches in organizations (AlKalbani et al. 2014; Ullah et al. 2013). The adoption of information security compliance is becoming increasingly the focus for adequately protecting organizational information (Boss and Kirsch 2007; Neubauer et al. 2006; Siponen et al. 2007; Von Solms 2001). Adopting the information security compliance approach, however, is both complex and challenging. This is because it involves (a) putting in place information security measures and mechanisms that can work together effectively, (b) satisfying the legal and security requirements of individual organizations and their stakeholders, and (c) maintaining both employees' and stockholders" confidence and trust (Attride-Stirling 2001; Neubauer et al. 2006; Steinbart et al. 2012). Often organizations introduce different methods to manage the information security compliance process. Some organizations, for example, introduce security trainings and awareness programs to provide their employees with the necessary skills and knowledge for responding to security threats (Bulgurcu et al. 2010). Other organizations use technical and administrative controls to guide employees to protect organizational information resources (Kolkowska and Dhillon 2012).

Several studies have investigated information security compliance in organizations. These studies can be categorized into two groups. The first group focuses on changing employees' attitude towards information security compliance in organizations (Bulgurcu et al. 2010; Herath and Rao 2009; Pahnila et al. 2007; Warkentin et al. 2011). In these studies, various theories such as the social cognitive theory (Bandura 1997), the social bond theory (Hirschi 1998), and the theory of protection motivation (Warkentin et al. 2011) are adopted for invistegating employees' interactions with information security mechanisms for information security compliance. This leads to the identification of various socio-organizational factors for affecting the adoption of information security compliance. Rewards for compliance, for example, have been found to have a significant impact on motivating employees' intention towards information security compliance (Herath and Rao 2009). The perceived risks and the consequences of non-compliance with information security policies have also been found as two key factors that encourage employees to be more proactive in undertaking higher information security precautions (Ryan 2004a; Ryan 2004b).

The second group concentrate on understanding employees' behavior towards information security compliance (Ifinedo 2013; Siponen et al. 2010; Son 2011; Vroom and Von Solms 2004). In these studies, several behavioral theories including the theory of planned behavior and reasoned action (Ajzen 1991), the deterrence theory (Straub Jr 1990), the threat avoidance (Liang and Xue 2010; Warkentin et al. 2011), and moral judgment (Myyry et al. 2009) have been explicitly or implicitly adopted for better understanding how security conscious behaviors are shaped in the adoption of information security compliance. The fear of sanction of non-compliance with information security policies, for example, has been found to have a significant impact on employees' behavior (Herath and Rao 2009; Kankanhalli et al. 2003). These studies have shown that understanding employee's attitudes and behavior towards information security compliance processes has a significant effect on increasing information security compliance in organizations.

Existing studies predominantly focus on individual's attitudes and behaviour in information security compliance. There is, however, more to be done on information security compliance with respect to the understanding of the complex socio-organizational dynamics associated with information security in organizations (Dhillon and Backhouse 2001; Vance et al. 2012). An investigation of such a dynamics leads to better understanding of the interactions among organizational, individual, and technical factors for shaping the adoption of information security compliance in organizations (Bulgurcu et al. 2010; Dhillon and Backhouse 2001). Investigating information security compliance at the organization level through empirical research based on different theoretical lenses would advance the current knowledge in the field (Vance et al. 2012). This study attempts to identify the most significant determinants of socio-organizational factors at the organization level for adopting information security compliance in organizations. Through the use of a quantitative analysis of the survey data, this study aims to examine the reliability and validity of the identified socio-organizational factors for information security compliance in organizations. These factors can then serve as a baseline for practices as well as the insights for the information security research that influence the adoption of information security compliance in organizations.

## 3　Framing Information Security Compliance in Theory

The aim of this research is to investigate the socio-organizational factors at the organization level for adopting information security compliance in organizations. It draws on a well-established theory including the technology-organization-environment (TOE) theory (Tornatzky et al. 1990) to examine the role of socio-organizational factors in the adoption of information security compliance in organizations. The TOE theory argues that the process by which technological innovations are adopted and implemented in organizations is conspired by the technological, organizational, and environmental contexts surrounding their operations (Tornatzky et al. 1990). Because this study focuses on investigating socio-organizational factors inside organizations, it considers only the technology and organization aspects of the TOE theory for adopting information security compliance in organization.

The technological context refers to the reliability of security technologies for satisfying information security policies and standards (Wimmer and Von Bredow 2002). Technology plays a critical role by providing organizations with secured transactions, protected access to information, and defence against hacker attacks (Venter and Eloff 2003). Adopting adequate security technologies capable of fulfilling the security requirements increases the trust and confidence of various stakeholders, leading to greater information security compliance (Moynihan 2004; Wimmer and Von Bredow 2002). Lambrinoudakis et al. (2003), for example, assess the security services offered by the public key infrastructure technology and audit for fulfilling the identified security requirements in an integrated e-government platform. Ebrahim and Irani (2005) and Lambrinoudakis et al. (2003) assert that technology capability could improve the normal functioning of information systems by reducing security risks and minimising cost impact in organizations.

Adopting adequate security technologies can help enforce policies, monitor and alert violations, and strengthen information protection for information security in organisations (Venter and Eloff 2003). Having the ability of security technologies to avoid operational systems incompatibility, such as the misfit between current work practices and security mechanisms is essential for the adoption of information security compliance (Smetters and Grinter 2002). Kaliontzoglou et al. (2005), for example, assess the effectiveness of different security technologies such as digital signature for enforcing security compliance. Ajzen (1991) explore the use of a role-based access control system for enforcing information security policies in organizations. Straub Jr (1990) examine the application of anti-spyware technologies for improving the information security compliance. These studies show the influence of security technologies for adopting information security policies in organizations.

The organizational context describes the organizational characteristics such as the organizational structure, communication processes and top management championship for promoting information security compliance. A well-developed set of organization initiatives on information security such as active commitment of top management directly affects employees behaviours in complying with information security standards and policies (Dhillon and Backhouse 2001; Sasse et al. 2001) and organizational processes in managing information security controls (Beautement et al. 2009). In fact, lack of management support in encouraging the adherence to information security policies has been singled out as a common reason for the weak implementation of information security policies in organizations (Knapp et al. 2006; Kolkowska and Dhillon 2012).

Promoting information security awareness in organizations can improve information security compliance in organizations (McIlwraith 2006). Having information security awareness programs in place is one way to improve information security compliance as such programs can raise users' knowledge and understanding of security policies and mechanisms in organizations (Puhakainen and Siponen 2010; Smith and Jamieson 2006). Bulgurcu et al. (2010), for example, point out that having information security awareness programs highly affects employee's beliefs about the benefit of compliance and the cost of non-compliance. Tsohou et al. (2008) show that the use of information security awareness and training programs can reduce the misuse of information security policies and procedures and increase users' avoidance of information security risks and threats in organizations.

Emphasising on individuals' roles and responsibilities towards information security is another way to promote information security in organizations (Herath and Rao 2009; Posthumus and Von Solms 2004). Employees with well-defined roles and responsibilities are more proactive in undertaking higher information security precautions (AlKalbani et al. 2015; Ryan 2004a). Herath and Rao (2009) state the importance of information security accountability in organization for information security compliance. Adams and Sasse (1999) point out that applying stipulated sanction for information security breaches in organizations encourages individuals to comply with information security standards and policies.

Developing appropriate operational processes to enhance information security in organizations can result in an efficient execution of information security controls (Knorr and Röhrig 2001). Information security processes and the way in which these processes are presented, integrated, and enforced are fundamental for effective information security compliance (Knorr and Röhrig 2001). Vroom and Von Solms (2004), for example, assert that compliance with information security policies can be improved if employees integrate information security mechanisms in their daily work practices. Backes et al. (2003) show that process integration is essential for developing appropriate operational processes that affect information security compliance in organizations.

Auditing and monitoring processes deal with the visibility of information security compliance in organizations (Neubauer et al. 2006). Developing appropriate auditing and monitoring processes is critical for information security compliance in organizations (Kolkowska and Dhillon 2012; Neubauer et al. 2006; Ransbotham and Mitra 2009; Steinbart et al. 2012). Auditing and monitoring processes, when appropriately enforced, could raise the speed of business operational execution and improve the overall effectiveness of information security mechanisms (Neubauer et al. 2006; Ransbotham and Mitra 2009). Steinbart et al. (2012) find that auditing and monitoring processes lead to increased acceptance of information security mechanisms. Kolkowska and Dhillon (2012) assert that auditing and monitoring processes could improve information security compliance in organizations.

The above discussion suggests that the adoption of information security compliance in organizations is influenced by the characteristics of technological and organizational contexts. This leads to the development of a conceptual model shown as in Figure 1 for the adoption of information security compliance for information security in organizations. The conceptual model hypotheses that technology capability, technology compatibility, management commitment, awareness and training, accountability, integration, and audit and monitoring will have a positive impact on the adoption of information security compliance in organizations at the organization level.

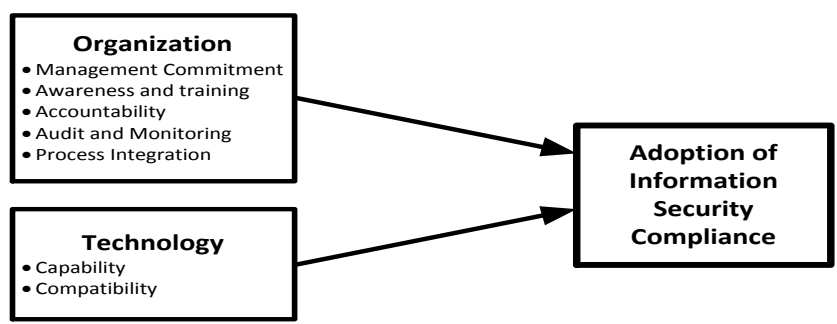

*Figure 1. A Research Model*

## 4   Research Methodology

The study aims to investigate the socio-organizational factors for affecting the information security compliance at the organization level in organizations. To facilitate the completion of this objective, a web-based survey and a paper-based survey are used for data collection. The construction of the survey is based on the previously validated questions to reflect the information security compliance context. The questionnaire consists of two parts. The first part is designed to capture participants' demographic information. The second part is intended to solicit participant's perceptions and opinions of the identified factors of information security compliance.

The measurement items used in this research are adopted from previous studies in information security compliance shown as in Table 2. A seven-point Likert scale is employed for each statement ranging from one describing "not important" until seven to indicate "highly important" (Miller 1987). Before administering the survey, the questionnaire is initially tested for content and constructs validity with experts in the field of information security and academics in information systems to ensure the semantics correspondence between the measurement items in the item pools and the underlying variables intended to be measured. Several of the original items are revised. The improved survey instrument based on the constructive comments from the expert review is used to conduct the survey.

The sample is comprised of public organizations in Oman in concurrence with the information security project run by the government of Oman across different ministries. The ample size is determined following the formula of Yamane (1973) with reference to prior studies in information security compliance. The participants are employees in public organizations in Oman. This respondent sample is selected because the participants have been exposed to an information security project within the government of Oman. Capturing the perception and opinion of these employees for adopting information security compliance provides reliable data for information security compliance in organizations.

The survey is conducted by hosting an online survey using the university Qualtrics application. Over a thousand invitations are sent to employees in public organisations in Oman via emails, messages in social media groups, and phone text messages with a link to the survey site. To boost the response rate, about 300 paper-based survey questionnaires are also randomly distributed to employees. Two months after issuing the invitation to participate in the online and paper-based survey, a follow-up phone call and a reminder email are again sent out to the targeted respondents. After five months (from July to December 2014), 326 responses are received (239 Online and 87 paper-based). After checking, 32 responses with missing data and aberrant response are excluded from the data file. This leads to a total of 294 completed questionnaires for the analysis.

The consideration of various types of organisations as well as participants working in different roles within the organisations ensures the robustness and generalizability of the research findings. The 294 responses represent employees from 64 organisations. Most participants' ages below or equals to 40 years. 51% of participants hold a 'bachelor's degree' in their education background. 40% of the respondents were female, whereas 60% of the respondents were male. The details of the sample demographics are reported in Table 1.

| **Profiles of Responding Participants** | **Frequency** | **Percentage** |
|---|---|---|
| Gender | | |
| -   Male | 175 | 60 |
| -   Female | 119 | 40 |
| Age | | |
| -   <=30 | 138 | 47 |
| -   31 – 40 | 133 | 45 |
| -   41 – 50 | 23 | 8 |
| -   51 – 60 | 0 | 0 |
| -   > 60 | 0 | 0 |
| Education Level | | |
| -   High School | 36 | 12 |
| -   Diploma/Advanced Diploma | 66 | 22 |
| -   Bachelor Degree | 149 | 51 |
| -   Master Degree | 38 | 13 |
| -   Doctoral Degree | 5 | 2 |
| Job Position Categories: | | |
| - Admission/Clerical | 114 | 39 |
| - Technician | 117 | 40 |
| - Managerial | 63 | 21 |
| Number of Years at current Role | | |
| - 1 - 3 | 107 | 36.4 |
| - 4 - 6 | 80 | 27.2 |
| - >= 7 | 107 | 36.4 |
| Organization Type | | |
| - Education | 52 | 18 |
| - Health Care | 34 | 12 |
| - ICT | 86 | 29 |
| - Trading | 43 | 15 |
| - Travel/Tourism | 15 | 5 |
| - Finance | 47 | 16 |
| - Agriculture | 17 | 6 |
| Total Number of Employees | | |
| - 1-50 | 3 | 1 |
| - 51 – 100 | 9 | 3 |
| - 101 – 250 | 31 | 11 |
| - 251 – 500 | 72 | 24 |
| - 501 – 1000 | 61 | 21 |
| - >1001 | 118 | 40 |

*Table 1. Summary of the participants' profiles*

## 5 Data Analysis Results and Research Findings

The constructs used in the research model are assessed based on (a) the reliability, (b) the convergent validity, (c) the discriminant validity, and (d) the adequacy of the model fit. To test the reliability, Cronbach's alpha is commonly used for testing the internal consistency of the identified factors by measuring the interrelatedness of the items in the survey questionnaire. For the data obtained from the survey, a reliability test is performed using SPSS 21.0 for Window based on the 294 responses. The Cronbach's alpha (α) as shown in Table 2 indicates that the average of the Cronbach's alpha value ranging from 0.75 to 0.82 which is considered as at a high level of reliability (Hair 2010). Based on these findings, the internal consistency of the survey instrument is acceptable and reliable.

An exploratory factor analysis is conducted to further examine the factor structure of the proposed research model by determining the convergent validity and the discriminate validity of the measurement model. The convergent validity test for a single factor is confirmed by examining both the average variance extracted (AVE) and the factor loadings of the indicators associated with each construct. The results indicate that two of the seven latent factors (Management Commitment, Awareness and training, Technology Compatibility, Technology capability, and Audit and Monitoring) have the AVE values equal to or exceeding the threshold value of 0.5, while two (Accountability and Process Integration) just marginally miss the 0.5 threshold value. In a strict sense, these two factors do not achieve the convergent validity (Fornell and Larcker 1981). However, the factor loadings of all seven latent factors range from 0.59 to 0.81 which are all statistically significant at $p = 0.05$. This shows that the presence of convergent validity in the model is supported (Bagozzi et al. 1991).

| Latent Factor | Range of Factor loading within constructs | AVE $\geq 0.5$ | Reliability $\alpha > 0.7$ | Items' Source |
|---|---|---|---|---|
| Management Commitment | 0.67 - 0.73 (4 Items) | 0.51 | 0.79 | Kenneth and Knapp 2006; Lee et al. 2004; Hayes et al. 1998 |
| Accountability | 0.59 - 0.81 (4 Items) | 0.49 | 0.79 | Chan, Woon and Kankanhalli 2005; Bulgurcu et al 2010; Kenneth Knapp 2006; Herath and Rao 2009a,b |
| Awareness and Training | 0.66 - 0.74 (3 Items) | 0.50 | 0.75 | Martins and Eloff 2001; Bulgurcu et al 2010; Siponen et al. 2010; Taylor and Todd, 1995 |

| Audit and Monitoring | 0.62 - 0.71 (4 Items) | 0.53 | 0.82 | Kenneth and Knapp, 2006; Herath and Rao 2009; |
| Process Integration | 0.67 – 0.71 (4 Items) | 0.48 | 0.78 | Chan, Woon and Kankanhalli 2005 |
| Technology Capability | 0.71 - 0.81 (3 Items) | 0.59 | 0.81 | Herath and Rao 2009; Siponen et al. 2010; Taylor and Todd 1995 |
| Technology Compatibility | 0.61 - 0.81 (3 Items) | 0.55 | 0.78 | Chan, Woon and Kankanhalli 2005; |

*Table 2. Summary of constructs' reliability and factor loadings*

The discriminant validity refers to which a construct is truly distinct from other constructs both in terms of how much it correlates with other constructs and how distinctly the measured variables represent only this single construct (Hair et al. 2010). The discriminant validity test of the single factor model is assessed by comparing the square root of the AVE for each construct against the inter-construct correlation estimates (Fornell and Larcker 1981). The result shows the square root of the AVE of each construct is higher than its correlation with other constructs (Management Commitment =0.711, Accountability =0.7, Awareness and training =0.707, Process Integration =0.7, Audit and Monitoring =0.728, Technology Capability =0.741, and Technology Capability =0.768). This shows that there are no cross-loadings for each item within these constructs. This result supports the discriminant validity of the measurement model (Fornell and Larcker 1981).

| Constructs | MC | Acc | ISA | ProInt | AudMon | TechCap | TechCom |
|---|---|---|---|---|---|---|---|
| **MC** | **0.711** | | | | | | |
| **Acc** | .648 | **0.700** | | | | | |
| **ISA** | .621 | .680 | **0.707** | | | | |
| **ProInt** | .605 | .632 | .617 | **0.700** | | | |
| **AudMon** | .578 | .642 | .579 | .661 | **0.728** | | |
| **TechCap** | .518 | .579 | .588 | .664 | .650 | **0.741** | |
| **TechCom** | .575 | .559 | .608 | .571 | .672 | .710 | **0.768** |

*Table 3. Summary of the constructs' discriminant validity*

The goodness-of-fit (GOF) measure is used to assess each single-factor model for their validity with various fitness indices, such as normed chi-square ($\chi^2$/d.f.), normed fit index (NFI), non-normed fit index (NNFI), comparative fit index (CFI), goodness of fit index (GFI), standardized root mean square residual (SRMR), and root mean-square error of approximation (RMSEA). Table 4 presents the GOF strength for each single-factor model indicating a good fit between variables in the dataset (Hu and Bentler 1999).

| Factor | No. of Items | x/df <3 | P >.05 | CFI >.95 | GFI >.95 | AGFI >.80 | SRMR <.09 | RMSEA <.05 | PCLOSE >.05 |
|---|---|---|---|---|---|---|---|---|---|
| MangCom | 4 | 0.655 | 0.519 | 1 | 0.998 | 0.989 | 0.0108 | 0.00 | 0.716 |
| Accont | 4 | 1.992 | 0.136 | 0.994 | 0.993 | 0.966 | 0.0211 | 0.058 | 0.330 |
| AwarTra | 3 | 0.014 | 0.907 | 1 | 1 | 1 | 0.0015 | 0.00 | 0.936 |
| ProcInt | 4 | 1.133 | 0.263 | 0.998 | 0.995 | 0.977 | 0.0198 | 0.034 | 0.489 |
| AuditMoni | 4 | 2.361 | 0.095 | 0.993 | 0.992 | 0.958 | 0.0196 | 0.068 | 0.263 |
| TechCap | 3 | 1.261 | 0.261 | 0.999 | 0.997 | 0.983 | 0.0111 | 0.030 | 0.419 |
| TechCom | 3 | 0.134 | 0.714 | 1 | 1 | 0.998 | 0.040 | 0.00 | 0.799 |

*Table 4. The GOF Results*

Based on the exploratory factor analysis and the confirmatory factor analysis for the research model, seven factors are identified. These are management commitment, accountability, awareness and training, operational process integration, audit and monitoring, technology compatibility, technology capability. This result reveals that these socio-organizational factors at the organization level have high level of reliability and validity for the adoption of information security compliance in organizations. This offers valuable insights on how information security compliance could be adopted in organizations.

The validated measurement model could be used in future research for developing and examining specific research hypothesis and theories relating to information security compliance in organizations. The results provide meaningful insights at organization level for adopting information security compliance in organizations. Furthermore, the results provide management and information security practitioners with a practical tool for early evaluation of the socio-organizational influences for the successful implementation of information security compliance in organizations by having early predictive measures in accepting information security practices in organizations.

This research contributes to the existing information security compliance literature in the following ways. First, the use of the TOE theory in this study extends the current understanding of information security compliance in terms of the value of socio-organizational aspects for information security compliance. Second, this study extends the current literature of information security compliance by investigating the factors at the organization level for adopting information security compliance, rather than predominantly focusing at the individual level using behavioral theories for changing employees' attitude and behaviors towards information security compliance.

# 6　Conclusion

This paper investigates the role of socio-organizational factors at the organization level for the adoption of information security compliance in organizations. It concludes that management commitment, accountability, awareness and training, process integration, audit and monitoring, technology capability, and technology compatibility, are significant factors for adopting information security compliance in organizations, particularly in public organizations in Oman. These socio-organizational factors offer valuable insights at the organizational level on how information security compliance could be achieved in organizations. This suggests that for shaping the adoption of information security compliance in organizations, it is necessary to go beyond users' attitude and behaviour.

While this study developed and tested the socio-organizational factors for influencing the adoption of information security compliance using the reliability test, exploratory factor analysis, and confirmatory factor analysis, it still has limitations that can be addressed in future. First, the generalizability of the research findings remains limited, since these findings have been validated in a single country. As a result, replicating this study in other countries, with different organizational and cultural settings would be a fruitful direction to gauge the generalizability of the study. Second, this study only collected the data from employees in public organizations. It has not surveyed other stakeholders, such as citizens and businesses that may have different perceptions for influencing the adoption of information security compliance in organizations. Finally, some other tangible measures of information security compliance could be considered. For instance, environmental measures could be investigated for improving information security compliance in organizations.